\documentstyle{elsart}
\input epsf

\begin{document}

\begin{frontmatter}

\title{Anomalous magnetic ordering in PrBa$_2$Cu$_3$O$_{7-y}$ single 
crystals:\\ Evidence for magnetic coupling between the Cu and
Pr sublattices}

\author[Troitsk,Dresden]{V.N. Narozhnyi\thanksref{CorrAuth}},
\author[Dresden]{D. Eckert},
\author[Dresden]{K.A. Nenkov\thanksref{OnLeave}},
\author[Dresden]{G. Fuchs},
\author[Moscow]{T.G. Uvarova},
\author[Dresden]{K.-H. M\"uller}

\address[Troitsk]{Institute for High Pressure Physics, Russian
  Academy of Sciences, Troitsk, Moscow Region, 142092, Russia}
\address[Dresden]{Institut f\"ur Festk\"orper- und Werkstofforschung 
  Dresden e.V., Postfach 270016, D-01171 Dresden, Germany} 
\address[Moscow]{Institute of Crystallography, Russian Academy of
Sciences, Leninski pr. 59, Moscow, 117333, Russia} 

\thanks[CorrAuth]{Corresponding author: Tel.: +7 095 334 08 08;
  Fax: +7 095 334 00 12; E-mail: narozh@ns.hppi.troitsk.ru}
\thanks[OnLeave]{On leave from International Laboratory of High 
  Magnetic Fields and Low Temperatures, Wroclaw, Poland}

\begin{abstract}

In Al-free $\rm PrBa_2Cu_3O_{7-y}$ single crystals the kink in the 
temperature dependence of magnetic susceptibility $\chi_{ab}(T)$, 
connected with Pr antiferromagnetic ordering, disappears after field 
cooling (FC) in a field $H\parallel ab$-plane. The kink in 
$\chi_c(T)$ remains unchanged after FC in $H\parallel c-$axis. As a 
possible explanation, freezing of the Cu magnetic moments, lying in 
the $ab-$plane, caused by FC in $H \parallel ab$, hinders their 
reorientation and, due to coupling between the Pr and Cu(2) 
sublattices, ordering of the $\rm Pr^{3+}$ moments. A field induced 
phase transition and a field dependence of the $\rm Pr^{3+}$ ordering 
temperature have been found for both $H\parallel c$ and $H\parallel 
ab$.
 

\end{abstract} 

 
\end{frontmatter}

\section{Introduction}

$\rm PrBa_2Cu_3O_{7-y}$ (Pr-123) is the anomalous member among the 
orthorhombic fully doped $R \rm Ba_2Cu_3O_{7-y}$ ($R=$Y, rare earth) 
cuprates (see Ref.\ \cite{Radousky,Hilscher,Merz} for a review). The 
transition temperature $T_{\mathrm{N}}$ for antiferromagnetic (AFM) 
ordering of the Pr magnetic moments, usually accepted as 17~K, 
\cite{Li,Kebede} is one or two orders of magnitude higher than 
exhibited by the other members of that series.  Moreover, magnetic 
ordering in the Cu(2) sublattice was found at $T \approx$~280~K even
though the compound is oxygen rich \cite{Cook}.  It was generally 
believed that Pr-123 is the only nonsuperconducting compound in that 
row, but very recently an indication on superconductivity was 
reported for Pr-123 grown by the traveling-solvent floating-zone 
method \cite{Ye}, see also Ref.\ \cite{Blackst}. This result is in 
sharp contrast to those obtained for crystals grown by the flux 
method (see Comment \cite{Narozh_Comm} on Ref.\ \cite{Ye}) and more 
work is necessary to clarify the situation.

Several models have been proposed to explain the absence of 
superconductivity in Pr-123 including (i) a valence of the Pr ion 
considerably greater than +3, (ii) magnetic pair breaking, (iii) Pr 
ions on Ba sites \cite{Blackst}, and (iv) hybridization of Pr and 
$\rm CuO_{2}$ layers (see, e.g., \cite{Radousky,Hilscher,Merz}). The 
situation with the Pr valence in Pr-123 is not completely clear up to 
now, see, e.g., Ref.\ \cite{Zhuo}, although strong evidence was 
obtained from inelastic neutron scattering \cite{Hilscher} and 
spectroscopic data \cite{Merz} for the predominance of +3 valence of 
Pr and Pr~4$f$--O~2$p_{\pi}$ hybridization. This hybridization 
should not only lead to the suppression of superconductivity but also 
to a considerable enhancement of the exchange interaction between the 
$\rm Pr^{3+}$ magnetic moments resulting in (i) a stabilization of 
the low-temperature non-zero value of these moments (which would 
vanish for dominating crystal field interactions) and (ii) an 
increase of $T_{\mathrm{N}}$. 

Although a $\lambda$-type anomaly at $T_{\mathrm{N}}$ was found from 
specific heat measurements \cite{Li,Kebede,Uma}, the magnetic 
susceptibility $\chi$ of Pr-123 continues to increase with decreasing 
temperature $T$ even below $T_{\mathrm{N}}$, 
\cite{Li,Kebede,Hilscher,Jayaram,Uma_x} contrary to the decrease of 
$\chi$ below $T_{\mathrm{N}}$ usually observed for AFM ordering.  
Only a weak change of the slope in the $\chi(T)$ dependence was found 
even for single crystals. In the first neutron diffraction study 
\cite{Li} a simple magnetic structure was proposed in which the Pr 
moments (0.74~$\mu_B$) point along the orthorhombic $c$ axis and 
alternate antiferromagnetically in all three crystallographic 
directions.  On the other hand, M\"{o}ssbauer spectroscopy 
\cite{Hodges} revealed AFM ordering with the moments tilted away from 
the $c$ axis by an angle $\theta$=65$^\circ$.  Tilting of magnetic 
moments has been also found by neutron diffraction \cite{Longmore}.  
All these results were questioned by the authors of NMR experiments 
\cite{Nehrke}, concluding that the Pr moment lies in the $ab$-plane 
and is only 0.017~$\mu_B$. This work was recently strongly criticized 
by Boothroyd $\em et~al.$, \cite{Boothroyd} who have investigated 
Al-free single crystals by neutron diffraction and found 
$\theta$=35$^\circ$ and $\mu_{\rm Pr}$=0.56~$\mu_B$. Moreover they 
observed, that the Pr ordering is accompanied by a counterrotation of 
the antiferromagnetic arrangement on the bilayer $\rm CuO_{2}$ planes 
about the $c$ axis with establishing of a noncollinear ordering of Cu 
moments below $T_{\mathrm{N}}$. It was underlined \cite{Boothroyd} 
that from symmetry considerations the coexistence of magnetic order 
on the $\em coupled\/$ Pr and Cu sublattices is not only consistent 
with, but even $\em requires\/$ a reorientation of the Cu magnetic 
structure during Pr magnetic ordering, see also Ref.\ 
\cite{Boothroyd_phb98}.

These reported results show, that the nature of magnetic ordering in 
Pr-123 is rather complicated. Previous magnetic measurements 
\cite{Jayaram,Uma_x} were performed on crystals grown in alumina 
crucibles, which leads to dissolving of Al in the sample. It is well 
known, that Al impurities can considerably influence the properties 
of cuprates (see, e.g., \cite{Longmore}). In the present work we have 
studied the magnetic properties of high quality Al-free Pr-123 single 
crystals. 
 
\section{Experimental}

Pr-123 single crystals were grown in Pt crucibles by the flux method 
\cite{Nar_jmmm}.  Electron-probe microanalysis has shown that the 
concentration of impurities in the crystals does not exceed the 
detection limit of the instrument used ($\sim $~1~at.~\%). Lower 
traces of impurities and especially of Pt can not be excluded. (The 
only known crucible's material that can give the ppm level of 
detrimental impurities in the flux grown Pr-123 crystals is 
unreactive BaZrO$_3$ \cite{Erb}). X-ray analysis has revealed single 
phase twinned orthorhombic material with lattice parameters 
$a$=3.868, $b$=3.911, and $c$=11.702~\AA. The crystals were annealed 
at $T$=450$^\circ$ in oxygen flow during one week. Two samples with 
the dimensions $\approx 1.2 \times 1 \times 0.2$~mm$^{3}$ and masses 
of 1.48 and 1.40~mg have been investigated.  Very similar results 
have been obtained for both crystals. The magnetization of the 
samples with field parallel ($M_c$) and perpendicular ($M_{ab}$) to 
the $c$-axis was measured at 1.7~K~$\leq T \leq $~300~K and 
$H\leq$~48~kOe by a Quantum Design SQUID magnetometer.  Magnetic 
susceptibility (determined as, e.g., $\chi_c(T)=M_c(T)/H$) has been 
measured typically at $H$=10~kOe.  

\section{Results and discussion}

\begin{figure} 
\epsfysize=4.7cm 
\centerline{\epsfbox{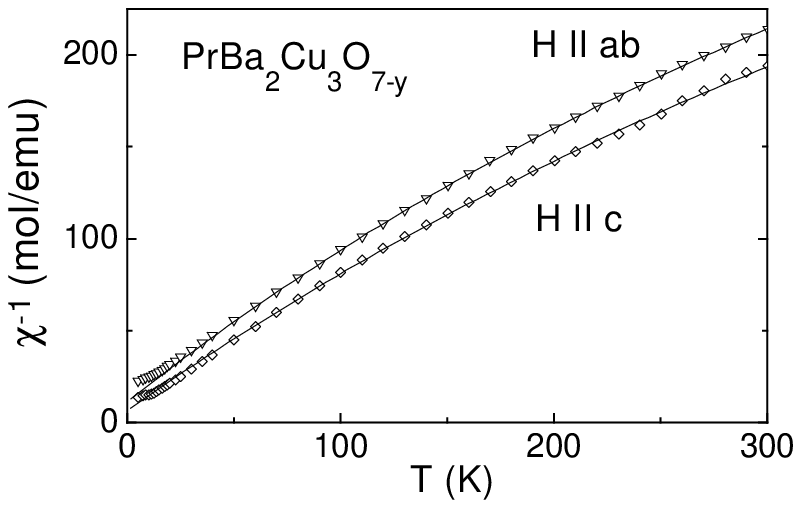}} 
\caption{$\chi^{-1}$ vs $T$ for Pr-123 single crystal for 
$H||ab$-plane and $H||c$-axis. The lines show the modified 
Curie-Weiss law fitted to the experimental data at 
50~K~$\leq T \leq$~300~K.} 
\label{fig1}
\end{figure}

The temperature dependence of the inverse magnetic susceptibility is 
shown in Fig.\ \ref{fig1} for the both directions of $H$.  The 
best fits of the data for 50~K~$\leq T \leq$~300~K to a modified 
Curie-Weiss law including a temperature independent $\chi_0$, see 
Ref.\ \cite{Hilscher}, are shown by lines in Fig.\ \ref{fig1}.  The 
values of the Pr effective paramagnetic moment obtained from this fit 
are 2.9 and 3.1~$\mu_B$ for $H||ab$ and $H||c$, respectively.  These 
values are in good agreement with previously published data for poly- 
and single crystals \cite{Li,Kebede,Hilscher,Jayaram,Uma_x} as well 
as with calculations of $\chi(T)$ based on the results of inelastic 
neutron scattering \cite{Hilscher} and taking into account the effect 
of crystalline electric field (CEF) splitting of the Pr$^{3+}$ free 
ion multiplet.  The obtained values of $\chi$(300~K) are also in 
agreement with these calculations for Pr$^{3+}$ and are approximately 
two times greater than the calculated value \cite{Hilscher} for 
Pr$^{4+}$. The observed sign of magnetic anisotropy 
($\chi_c>\chi_{ab}$) corresponds to the schemes of CEF splitting of 
the low lying quasitriplet (well separated from higher levels of 
Pr$^{3+}$) proposed in Refs.\ \cite{Hilscher} and \cite{Boothroyd_x}. 
The other sign of anisotropy ($\chi_c<\chi_{ab}$) was proposed by 
assuming another scheme of levels \cite{Soderholm}. This result is 
not in agreement with our data. It should be noted, that the 
conclusions from NMR experiments, \cite{Nehrke} i.e., 
$\chi_c<\chi_{ab}$, a very small value of $\mu_{\rm Pr}$, and 
temperature independence of $\chi$ at $T<$ 20~K are in strong 
disagreement with our results. So, our data can be considered as a 
further evidence for the $3+$ valence of Pr in Pr-123 with the scheme 
of levels proposed in Refs.\ \cite{Hilscher} or \cite{Boothroyd_x}.

\begin{figure} 
\epsfysize=9.5cm 
\centerline{\epsfbox{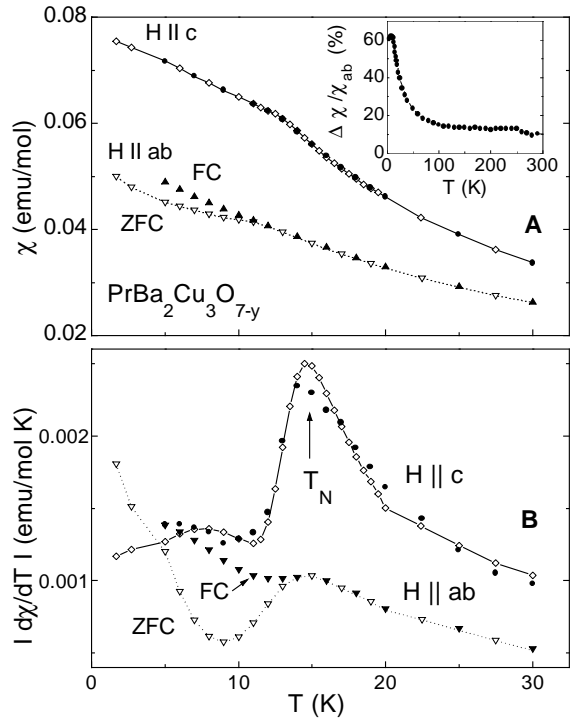}}
\caption{(A) $\chi$ vs $T$ and (B) the absolute value of the
susceptibility derivative $|\d \chi(T)/\d T|$ vs $T$ for a
Pr-123 single crystal for two directions of $H$.  
Solid and dotted lines connecting ZFC data (open symbols) are
guides for the eye. Solid symbols in both parts of the figure 
represent the data measured at $H$=20~kOe after FC in $H$=20~kOe 
($H\parallel$ $ab$-plane) and in $H$=48~kOe ($H\parallel$ $c$-axis). 
The inset shows the anisotropy parameter 
$(\chi_c-\chi_{ab})/\chi_{ab}$ vs $T$.} 
\label{fig2} 
\end{figure}

The anisotropy of $\chi$ is clearly seen in Figs.\ \ref{fig1} and 
\ref{fig2}A. The degree of magnetic anisotropy 
$\Delta\chi/\chi_{ab}=(\chi_c-\chi_{ab})/\chi_{ab}$ is $\approx$10\% 
at $T$=300~K and increases with decreasing temperature to the value
$\approx$60\% at $T$=15~K (see the inset of Fig.\ \ref{fig2}A). This 
value is considerably larger, than that reported for crystals grown 
in alumina crucibles \cite{Jayaram,Uma_x} ($\approx$10\% at $T$=5~K). 

In Fig.\ \ref{fig2}A distinct kinks in $\chi(T)$ are clearly visible 
for the both field directions. Earlier, the anomaly at 
$T_{\mathrm{N}}$ for oxygen rich Pr-123 could be seen only in the 
temperature dependence of the derivative \cite{Jayaram} $\d 
\chi(T)/\d T$ and in the anisotropy of $\chi(T)$ \cite{Uma_x} but not 
for the $\chi(T)$ dependence itself. (It should be pointed out, that 
the corresponding anomaly in $\chi(T)$ at $T_{\mathrm{N}}$ was 
observed for oxygen depleted Pr-123 crystals \cite{Uma_x}.) The 
maxima in the $|\d \chi/\d T|$ vs $T$ dependencies marked by the 
arrow in Fig.\ \ref{fig2}B are located at $T \approx$~15~K. In 
accordance with \cite{Li,Kebede} we consider the position of the 
maximum as $T_{\mathrm{N}}$.  This value is slightly lower than the 
usually cited \cite{Li} value of $T_{\mathrm{N}}$=17~K.  However, 
depending on sample preparation techniques and methods of 
measurements, the value of $T_{\mathrm{N}}$ varies from 11 to 20~K 
\cite{Li,Kebede,Hilscher,Uma_x,Longmore,Boothroyd} even for oxygen 
rich compounds. For oxygen reduced samples values of $T_{\mathrm{N}} 
\leq$~11~K were reported \cite{Uma_x,Hodges,Longmore,Kebede_jap}. It 
is well known, that it is difficult to oxidize the cuprates single 
crystals.  Therefore, oxygen deficiency could be a reason for the 
lower value of $T_{\mathrm{N}}$ for our crystals.

The data obtained for the two directions of $H$ after zero field 
cooling ZFC have been marked by the open symbols connected by lines 
in both parts of Fig.\ \ref{fig2}. Unexpectedly, field cooling FC 
from $T$=40~K in a field of $H$=20~kOe for $H||ab$-plane suppresses 
fully the anomaly in $\chi_{ab}(T)$ (solid symbols in lower curves in 
Figs.\ \ref{fig2}A and B). At the same time FC for $H||c$-axis even 
in a field of $H$=48~kOe has no influence on $\chi_c(T)$ and 
$|\d \chi_{ab}(T)/\d T|$ curves (solid symbols in upper curves). For 
$H||ab$-plane the influence of FC in one order of magnitude 
smaller field, $H$=5~kOe, on $|\d \chi_{ab}(T)/\d T|$ curve was 
observed (not shown here). After ZFC the field itself does not 
destroy the kink in $\chi(T)$.  Only a decrease of the peak in the 
$|\d \chi_c(T)/\d T|$ curve was observed by increasing the field. 
This decrease is more pronounced for $H||ab$-plane.  It should be 
underlined, that FC has not only an effect on $\chi(T)$ curves but, 
in the case of $H||ab$-plane, it $\em nearly~completely~suppresses\/$ 
the anomaly in the $\chi_{ab}(T)$ dependence suggesting the 
suppression of AFM ordering of the Pr magnetic moments by FC in 
$H||ab$-plane. 

\begin{figure} 
\epsfysize=4.9cm 
\centerline{\epsfbox{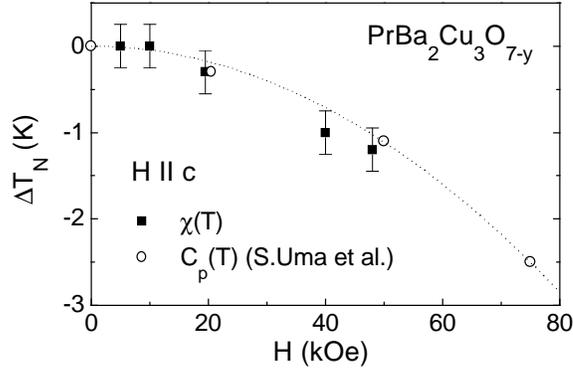}} 
\caption{Decrease of $T_{\mathrm{N}}$ of Pr-123 single crystals for 
$H\parallel$ $c$-axis. Open symbols are the results of specific heat 
measurements of Uma {\it et~al}. \protect\cite{Uma}. The dotted line
shows the best fit of Uma's data to a 
$\Delta T_{\mathrm{N}} \sim H^2$ dependence.} 
\label{fig3} 
\end{figure}

The distinct anomaly in the $\chi(T)$ dependence for our high quality 
single crystals makes it possible to see clearly a decrease of 
$T_{\mathrm{N}}$ with increasing field, see Fig.\ \ref{fig3}. 
Earlier, a shift 
$\Delta T_{\mathrm{N}}=T_{\mathrm{N}}(H)-T_{\mathrm{N}}(0)$ was not 
observed in $\chi(T)$ experiments in fields up to 90~kOe 
\cite{Kebede,Hilscher} contrary to the results of specific heat 
measurements \cite{Hilscher,Uma} which clearly indicated such a 
shift.  Our results are in good agreement with the specific heat data 
of Uma $\em et~al.$, \cite{Uma} obtained on crystals with 
$T_{\mathrm{N}}(0)$=16.6~K, and with the proposed by them $\Delta 
T_{\mathrm{N}} \sim H^2$ dependence.

\begin{figure} 
\epsfysize=9.2cm
\centerline{\epsfbox{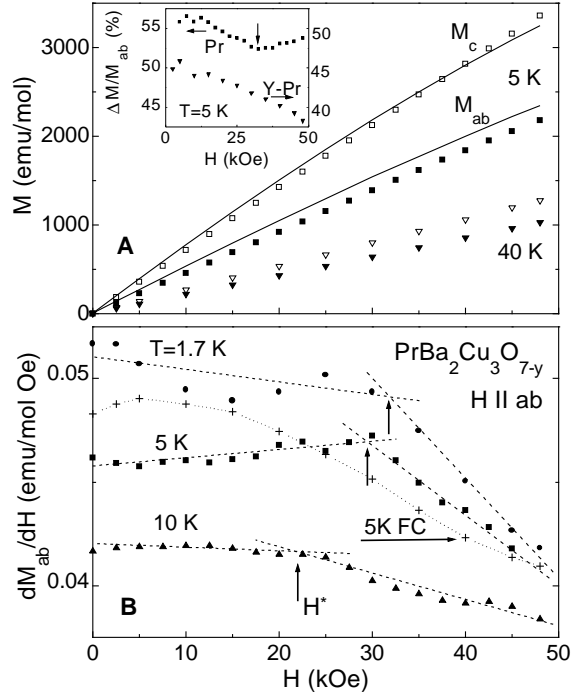}}
\caption{(A) Isothermal magnetization of a Pr-123 single crystal vs 
$H$ for the two field directions at $T$=5~K and 40~K. Solid lines 
show the magnetization (in emu/mol~Pr) of a $\rm 
Y_{0.4}Pr_{0.6}Ba_2Cu_3O_{7-y}$ (YPr-123) single crystal.  (B) Field 
derivative of the in-plane magnetization $\d M_{ab}(H)/\d H$ vs $H$ 
for Pr-123 single crystal at T=1.7, 5, and 10~K after ZFC. Lines
represent the linear fits to the low and high field parts of the 
data. Vertical arrows mark the transition fields $H^*$.  Crosses 
connected by dotted lines show the results obtained at $T$=5~K after 
FC in $H$=20~kOe. The inset shows the normalized anisotropy in
magnetization $(M_c-M_{ab})/M_{ab}$ vs $H$ at $T$=5~K for Pr-123 and 
YPr-123 single crystals obtained after ZFC.} 
\label{fig4} 
\end{figure}

We have also studied the isothermal magnetization of the Pr-123 
crystals at 1.7~K~$\leq T \leq$~40~K. Some of the obtained results 
are shown in Fig.\ \ref{fig4}A for the both directions of field.
At all temperatures the $M(H)$ dependencies are nearly linear and the 
magnetic anisotropy is more pronounced at lower temperatures. For 
comparison the magnetization curves of a 
$\rm Y_{0.4}Pr_{0.6}Ba_2Cu_3O_{7-y}$ (YPr-123) single crystal 
\cite{Y-Pr} grown under the same conditions are also shown for 
$T$=5~K. The $M_c(H)$ curve for YPr-123 is close to the corresponding 
$M_c(H)$ curve for Pr-123, whereas the $M_{ab}(H)$ curve for YPr-123 
lies somewhat higher giving a smaller degree of the magnetic 
anisotropy.  The stronger magnetic anisotropy for the pure Pr-123 may 
be caused by some anisotropic (e.g.  pseudodipolar) type of 
interaction between the $\rm Pr^{3+}$ ions because the crystal fields 
acting on $\rm Pr^{3+}$  should not be affected by the dilution of Pr 
by Y.

Analyzing the derivative of $M_{ab}(H)$ we observed a change in the 
behavior at certain transition fields $H^*$ marked in Fig.\ 
\ref{fig4}B by vertical arrows. Whereas $\d M_{ab}(H)/\d H$ is almost 
constant at low fields, $H<H^*$, it decreases for $H>H^*$. For the 
sake of simplicity the low and high field parts of the data were 
approximated by straight lines as shown on Fig.\ \ref{fig4}B.  
Careful analysis of $\d M_{ab}(H)/\d H$ curves at $T$=1.7~K, 2.7~K 
(not shown in the picture), and 5~K has revealed the tendency of 
$\d M_{ab}(H)/\d H$ to increase with increasing field just below
$H^*$.  A more rapid increase of $M$ and hence an increase of $\d 
M(H)/\d H$ below the transition field is a characteristic feature of 
field induced phase transitions (FIPT) \cite{Zvezdin}. The observed 
behavior can be considered as an indication on a weak FIPT in 
Pr-123 at transition field $H^*$ (probably of spin-reorientation 
type). The exact nature of this transition has to be settled.  

Similarly as the anomaly in the $\chi_{ab}(T)$ dependence this FIPT 
can be suppressed by FC in $H||ab$-plane. It can be seen from the 
$\d M_{ab}(H)/\d H$ dependence obtained at $T$=5~K after FC and shown 
on Fig.\ \ref{fig4}B by crosses connected by dotted lines.  Obviously 
there is no indication of a FIPT for this curve.  The FC 
$\d M_{ab}(H)/\d H$ curve for Pr-123 measured at $T$=5~K is similar 
to the corresponding ZFC curve for the YPr-123 crystal (not shown in 
the picture), for which no anomaly in $\chi(T)$ was observed even 
after ZFC. 

It should be also noted, that the field dependence of the anisotropy 
parameter $\Delta M/M_{ab}=(M_c(H)-M_{ab}(H))/M_{ab}(H)$ at $T$=5~K 
has a clearly visible minimum at $H \approx$~30~kOe (marked by 
vertical arrow in the inset of Fig.\ \ref{fig4}A), that is very 
close to $H^*$(5~K) obtained from the $\d M_{ab}(H)/\d H$ curve shown 
in Fig.\ \ref{fig4}B. This minimum can be considered as another
manifestation of FIPT in Pr-123. On the other hand, the 
$\Delta M/M_{ab}$ vs $H$ curve for YPr-123 (see the inset of 
Fig.\ \ref{fig4}A) does not indicate any transition.  

\begin{figure} 
\epsfysize=5.0cm 
\centerline{\epsfbox{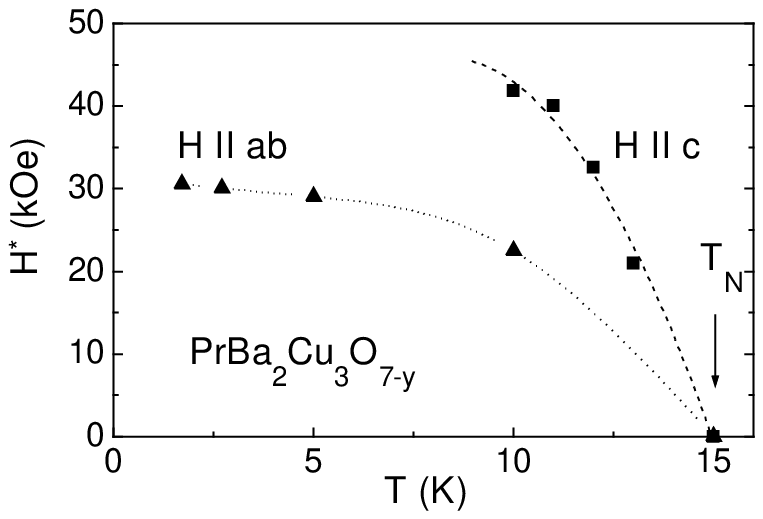}} 
\caption{Transition field $H^*$ (determined as shown by arrows in 
Fig.\ 4B) vs $T$ for two directions of $H$. Lines are guides for 
eye.} \label{fig5} \end{figure}

The FIPT was observed also for $H||c$-axis, the value of $H^*(T)$ 
being about two times higher, than for $H||ab$-plane. A magnetic 
phase diagram for the FIPTs in Pr-123 is shown in Fig.\ \ref{fig5}.  
The estimation of $H^*(0)$ for $H||ab$-plane gives 31~kOe.  The 
observed FIPTs are very weak. Therefore, a more complicated phase 
diagram can not be excluded.

For the interpretation of the observed results the mentioned above 
model \cite{Boothroyd} of the coupled Cu and Pr sublattices can be 
used. In that model the Pr ordering is accompanied by the 
reorientation of the Cu magnetic structure due to the Pr-Cu 
magnetic interaction. (Very recently it was found \cite{Hill}, that 
Pr sublattice in fact orders in a long period incommensurate 
structure just below $T_{\mathrm{N}}$ and reorders to commensurate 
AFM structure at lower temperatures.) Since the influence of FC has 
been observed by us only for $H||ab$-plane, it should be connected 
with Cu moments, lying in the $ab$-plane \cite{Longmore,Boothroyd}.  
At the same time, the weak FIPTs observed at $T<T_{\mathrm{N}}$ for 
both directions of $H$ are connected most probably with the 
reorientation transitions of Pr moments, tilted away from the $c$ 
axis \cite{Hodges,Longmore,Boothroyd}.  The mechanism of the proposed 
suppression of $\rm Pr^{3+}$ AFM ordering by FC in $H||ab$-plane may 
be as follows.  During FC the Cu moments could be frozen by 
sufficiently strong $H||ab$-plane.  Freezing of Cu magnetic moments 
hinders the AFM ordering in the Pr sublattice due to coupling with 
the magnetic Cu sublattice. There are some indications on frustration 
in the Cu magnetic structure as well as on short range ordering in 
the Cu sublattice \cite{Boothroyd}.  Therefore, it may be favorable 
for the Cu moments to freeze during FC in $H||ab$-plane.  So, the 
observed behaviour can be considered as an evidence for the existence 
of considerable coupling between the Cu and Pr magnetic sublattices 
in Pr-123.

\section{Conclusion}

We have observed for Pr-123 single crystals a large magnetic 
anisotropy and a kink in the $\chi(T)$ dependence, connected with Pr 
AFM ordering. A practically complete suppression of the kink after FC 
in $H||ab$ was found. At the same time this anomaly remained 
uninfluenced by FC in $H||c$. A possible explanation of these 
observations is connected with the magnetic coupling between the Cu 
and Pr sublattices and the suppression of AFM ordering in the Pr 
sublattice by freezing of Cu moments, lying in the $ab$-plane, caused 
by FC in $H||ab$.  Weak FIPTs found for both directions of $H$ are 
connected most probably with a Pr$^{3+}$ spin-reorientation.

\ack We thank S.-L. Drechsler and M. Wolf for discussions. This work 
was supported by RFBR grants 96-02-00046G, 96-02-16305a; DFG grant 
MU1015/4-1.

\newpage
  
\newpage
 
\newpage
 
\newpage

\newpage
 
\begin{thebibliography}{99} 

\bibitem{Radousky}H. B. Radousky, J. Mater. Res. 7 (1992) 1917.

\bibitem{Hilscher}G. Hilscher $\em et~al.$, Phys. Rev. B 49 (1994) 
535.

\bibitem{Merz}M. Merz $\em et~al.$, Phys. Rev. B 55 (1997) 9160.

\bibitem{Li}W-H. Li $\em et~al.$, Phys. Rev. B 40 (1989) R5300.
 
\bibitem{Kebede}A. Kebede $\em et~al.$, Phys. Rev. B 40 (1989) 4453.
 
\bibitem{Cook}D. W. Cook  $\em et~al.$, Phys. Rev. B 41 (1990) R4801.

\bibitem{Ye} Z.  Zou, J.  Ye, K.  Oka, and Y.  Nishihara, Phys.  Rev.  
Lett. 80 (1998) 1074. 
  
\bibitem{Blackst}H. A. Blackstead $\em et~al.$, Phys. Rev. B 54 
(1996) 6122.

\bibitem{Narozh_Comm}V. N. Narozhnyi and S.-L. Drechsler, 
Phys. Rev. Lett. (to be published 14 December 1998).

\bibitem{Zhuo}Y. Zhuo $\em et~al.$, Phys. Rev. B 56 (1997) 8381.

\bibitem{Uma}S. Uma $\em et~al.$, J. Appl. Phys. 81 (1997) 4227.

\bibitem{Jayaram}B. Jayaram  $\em et~al.$, Phys. Rev. B 52 (1995) 89.

\bibitem{Uma_x}S. Uma $\em et~al.$, Phys. Rev. B 53 (1996) 6829. 
 
\bibitem{Hodges}J. A. Hodges $\em et~al.$, Physica C 218 (1993) 283.

\bibitem{Longmore} A. Longmore $\em et~al.$, Phys. Rev. B 53 (1996) 
9382.

\bibitem{Nehrke} K. Nehrke and M. W. Pieper, Phys. Rev. Lett. 
76 (1996) 1936.
  
\bibitem{Boothroyd}A. T. Boothroyd $\em et~al.$, Phys. Rev. Lett. 
78 (1997) 130.

\bibitem{Boothroyd_phb98}A. T. Boothroyd, Physica B 241-243 (1998) 792.

\bibitem{Nar_jmmm}V. N. Narozhnyi and T. G. Uvarova, J. Magn. Magn. 
Mater. 157/158 (1996) 675.

\bibitem{Erb}A. Erb, E. Walker, and R. Fl\"ukiger, Physica C 258 
(1996) 9.

\bibitem{Boothroyd_x}A. T. Boothroyd, S. M. Doyle, and R. Osborn, 
Physica C 217 (1993) 425.

\bibitem{Soderholm} L. Soderholm, C.-K. Loong, G. L. Goodman, and
B. D. Dabrowski, Phys. Rev. B 43 (1991) 7923.
  
\bibitem{Kebede_jap}A. Kebede $\em et~al.$, J. Appl. Phys. 69 (1991) 
5376.
  
\bibitem{Y-Pr}Complete results of investigation of YPr-123 as well
as GdPr-123 crystals will be published elsewhere.

\bibitem{Zvezdin} A. K. Zvezdin, in $\em Handbook~of~magnetic~
materials$, edited by K. H. J. Buschow, Elsevier, Amsterdam, 1995.

\bibitem{Hill} J.P. Hill $\em et~al.$, Phys. Rev. B 58 (1998) 11211.
  
\end{thebibliography}
\end{document}